# Mission Apollo: Landing Optical Circuit Switching at Datacenter Scale

Ryohei Urata, Hong Liu, Kevin Yasumura, Erji Mao, Jill Berger, Xiang Zhou, Cedric Lam, Roy Bannon, Darren Hutchinson, Daniel Nelson, Leon Poutievski, Arjun Singh, Joon Ong, Amin Vahdat

Google LLC

## ABSTRACT

In this paper, we describe Apollo, to the best of our knowledge, the world's first large-scale production deployment of optical circuit switches (OCSes) for datacenter networking. We will first describe the infrastructure challenges and use cases that motivated optical switching inside datacenters. We then delve into the requirements of OCSes for datacenter applications: balancing cost, port count, switching time, and optical performance, which drive design choices and implementation details of our internally developed 3D MEMS-based OCS. To enable the Apollo optical switching layer, we employ circulators to realize bidirectional links through the OCS, effectively doubling the OCS radix. The OCS and circulator design choices were critical for meeting network bandwidth, scale, and cost targets. We review the critical co-design of WDM transceiver technology for these OCS plus circulator-based bidirectional links and their corresponding physical impairments, delivered over four generations/speeds of optical interconnect. Finally, we conclude with thoughts on future directions in hardware development and associated applications.

## 1. Introduction

Over the past few decades, large hyperscale datacenters [1] have enabled new, global-scale Internet services, from the proliferation of web search, e-commerce, social media platforms, to the public cloud. In recent years, machine learning (ML) applications and workloads have further amplified the importance of such large-scale compute capability to open new frontiers in a variety of fields, from biology and medicine [2], mastering the game of Go [3], to enhancing the aforementioned Internet services [4, 5].

Datacenter networking provides the interconnectivity and scale needed for executing these functions and services. In previous work [6], we outlined the evolution and development of our datacenter network, from the initial use of vendor-based gear to the creation of a home-grown Clos-fabric-based network [7] achieving unprecedented scale, with similar efforts across the industry [8-11].

While industrial best practices have focused on pure packet solutions employing Clos topologies as the basis for large-scale datacenter networks, the research community has developed a number of potentially game changing network designs around the premise of incorporating optical circuit switches (OCSes) into the datacenter architecture [12-20] by taking advantage of the dynamism and flexibility that electrical packet switch (EPS)-only networks lack. OCS technologies hold a number of benefits relative to EPSes, being data rate and wavelength agnostic, low latency, and extremely energy efficient. In particular, by simply steering light from source to destination with no intermediate processing, the OCS is agnostic/transparent to bit rate, protocol, framing, line coding, and modulation formats, thus allowing the optical transceiver interfaces to evolve independently.

At the top level, the benefits of OCSes for datacenter networking include:

- Supporting a range of novel network topologies best suited to the communication patterns of individual services [16].
- Removing a subset of EPSes from otherwise static topologies for reduced power consumption, lower latency, and lower cost [12-18].
- Incremental evolution of building-scale networks across multiple generations without incurring the high cost and higher service inconvenience of vacating a building's worth of capacity for potentially months to first tear down the previous network fabric and then bring up a new fabric.
- Vertical integration with bandwidth-intensive and latency-sensitive applications such as ML training [19], with predictable, repeating communication patterns that shift intensive communication among changing subsets of computation/accelerator nodes.

Each of the above use cases in isolation promise integer factor improvements in cost, latency, and bandwidth efficiency of the network through clever integration of optical switching technology into electrical packet switched networks. These opportunities are one of the most compelling architectural opportunities in datacenter networking. While each requires substantial control software evolution, the primary impediment to delivering on these opportunities has been the lack of a manufacturable, cost-efficient, and reliable optical circuit switching hardware to serve as the basis for development and deployment of next generation datacenter technologies. At least some conventional wisdom says that producing such hardware is impractical; an explicit goal of this paper is to re-examine and perhaps to reset such conventional wisdom.

In this paper, we present the design and implementation of Apollo, which we believe is the first large-scale deployment of optical circuit switching for datacenter networking. The Apollo OCS platform consists of a home-grown, internally developed OCS (Palomar), circulators, and customized wavelength-division-multiplexed (WDM) optical transceiver technology supporting bidirectional links through the OCS and circulators. Apollo serves as the backbone of all of our datacenter networks, having been in production for nearly a decade in support of all of the datacenter use cases outlined above. In separate work, we describe how Apollo delivers substantial improvements to the evolution, reliability, cost efficiency, power consumption, and application performance of our infrastructure [21]. However, none of these efforts would have been possible without the presence of a reliable, cost effective OCS and corresponding optical interconnect.

To realize the above at datacenter scale, our top-level design principles are as follows: 1) Palomar design emphasizes manufacturability, serviceability, and reliability. Image processing of a single camera image for Micro-Electro-Mechanical Systems (MEMS) mirror control greatly simplifies manufacturing. The front and back chassis architecture allows the optical core and primary printed circuit board assemblies (PCBAs) to be assembled and tested separately prior to full product integration. This architecture enhances testability, yield, and also enables field serviceability (field replaceable PCBAs). With these principles, we have created a highly reliable and manufacturable 136x136 OCS, with millisecond-scale switching time and worst-case insertion loss of 2dB and return loss of -38dB. While we cannot reveal specific cost figures, Palomar is lower in per port cost than equivalent throughput EPSes. 2) Circulators further enhance this cost advantage by doubling the OCS effective port count. 3) Equally critical has been the co-design of the optical transceivers for the OCS-based links. Leading the industry, we have delivered robust, low cost WDM transceivers over four generations of optical interconnect speeds (40, 100, 200, 400GbE) through a combination of high-speed optics, electronics, and signal processing technology development.

The remainder of this paper is organized as follows. In Section 2, we discuss multiple abstract use cases enabled by Apollo. In addition to traditional EPSes, the networking fabric consists of critical hardware elements of the OCS, WDM optical transceiver technology, and optical circulators. In Section 3, we review the challenges and requirements of OCSes for datacenter applications, which drive the corresponding design implemented for our internally developed MEMS-based OCS: Palomar. In Section 4, we also review the critical co-design of WDM transceiver technology for the OCS plus circulator-based bidirectional links and their corresponding physical impairments. The Apollo fiber infrastructure and physical layout are reviewed in Section 5. We conclude with thoughts on future directions in hardware development and the associated new applications.

## 2. Use Cases

Below, we review two use cases for Apollo: 2.1) Datacenter network evolution through replacement of EPS with OCS. 2.2) Flexible network topology configuration in support of ML training.

### 2.1 Datacenter Network

Figure 1a) illustrates a traditional datacenter network architecture, with Spine blocks connecting Aggregation blocks (ABs) [6]. Interconnection of top-of-rack (TOR) switches to the AB is implemented with parallel-fiber-based optical transceivers (either short reach (SR) multi mode or parallel single mode (PSM)) and corresponding fibers. Figure 1b) illustrates the evolved datacenter network architecture incorporating the Apollo OCS layer. The Apollo layer replaces the Spine blocks for significant cost and power savings through elimination of the electrical switches and optical interfaces that are used to implement the Spine layer.

We employ single mode WDM optical transceivers for the inter-AB connections. WDM optics maximize the efficiency and usage of OCS ports, as opposed to a PSM solution. Single mode operation is needed for compatibility and scaling of the OCS technology and is capable of supporting a large and ever growing network infrastructure, made more difficult as per-lane bandwidths increase. Furthermore, circulators are coupled to the optical transceivers to operate these single mode optical links in a bidirectional manner so that full duplex communication is achieved for each single strand of fiber and each OCS port. This halves the required number of fibers and OCS ports. With the OCS, circulator, and fiber components being largely data rate agnostic, they can be used across multiple generations of networking and interconnect of different speeds. We will discuss the implementation details of the OCS, WDM transceivers, and circulators in Section 4.

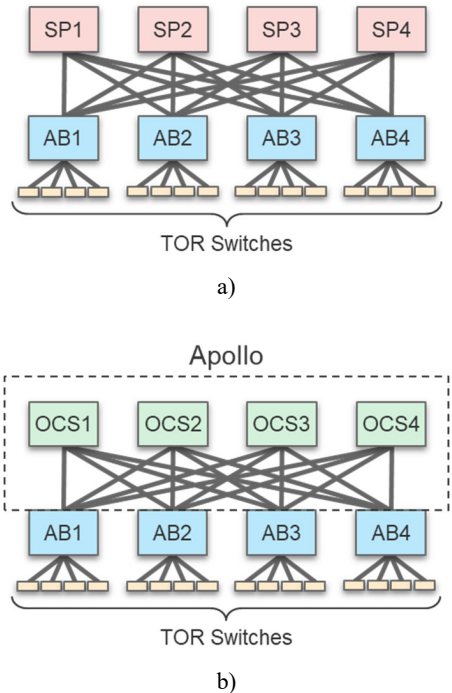

Figure 1: Evolution of Network Architecture. a) Traditional hierarchical datacenter network architecture with Spine blocks connecting Aggregation blocks. b) Apollo OCS-based datacenter network architecture. Spine blocks are replaced by cut-through OCSes to eliminate the Spine.

The OCS-based network adds a large amount of flexibility to a normally static Clos-based networking fabric. Although we could consider insertion of the OCS layer in different locations of the network (parallel to Spine blocks, in between TOR switches and AB, in parallel with AB), we focus on the inter-AB option here considering the combination of cost savings and ease of adoption (due to traffic locality within an AB).

*2.1.1 Topology Engineering*
When moving towards spine-free architectures, with the elimination of routing capability by the spine, a degree of fast/dynamic network topology reconfiguration is desirable in addition to regular electrical-switch-level traffic engineering (i.e., equal-cost-multi-path (ECMP) or weighted-cost-multi-path (WCMP) routing). As an example, this topology engineering capability allows reconfiguration of the fabric connectivity to target maximization of inter-AB bandwidth in the event of an increase in long-lived traffic demand (i.e., elephant flows) between two particular ABs or multiple ABs. Topology engineering is thus used to achieve equivalent network throughput with fewer links (higher efficiency) or increased throughput with the same number of links (higher performance). Additional details can be found in [21].

*2.1.2 Expansion*
With expansion capability, the size of the network fabric can be augmented incrementally, allowing an initial number of ABs to be deployed with additional ABs added to the fabric as needed (pay as you grow). The Apollo layer achieves the required interconnect re-striping (configure from old to new connectivity) across ABs in an automated fashion (Figure 2). Re-striping of the network to support the new configuration could have been done through patch panels. However, due to the sheer size of the fabric and corresponding interconnect number, the time needed to rewire and coordinate requalification of links becomes unmanageable at scale.

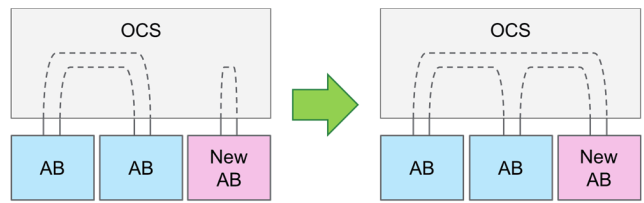

Figure 2: Simplified illustration of fabric expansion. OCS configuration re-stripes the network connectivity according to the number of ABs used.

For each expansion, the appropriate links are drained, reconfigured with the OCS, then qualified before releasing to production traffic flows. For the last step, a link qualification process would consist of the following: a cable audit is done to first check baseline packet transmission, followed by a bit-error-rate test (BERT) to check and verify the quality of the link.

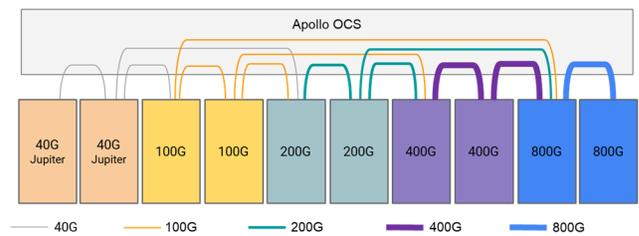

Figure 3: Interoperability within the Apollo-based network. Switch ASIC and optics technology of a given generation and speed compose each AB. Five generations and speeds of AB are shown above. Backward compatibility of the WDM optics at the top of each AB (which transit the circulators and OCS) allows interoperability from AB to AB.

*2.1.3 Rapid Tech Refresh*
The above expansion capability leads to the possibility of mixing ABs based on different speed/bandwidth generations of optical transceivers and switch ASICs. Interoperability between heterogeneous ABs is ensured through the compatibility of optical transceiver specifications across multiple generations, as illustrated in Figure 3 (implementation detail covered in Section 4). This enables faster introduction of new technology, as well as incremental tech refresh. The OCS-based expansion and interconnect interoperability thus enable optimally sized datacenter networks incorporating the latest and lowest cost networking technologies (i.e., optical interconnect and switch ASIC).

## 2.2 Machine Learning

Large ML training models are communication intensive. Cloud TPU v4 Pod connects 4,096 TPU v4 chips by an ultra-fast interconnect that provides 10x the bandwidth per chip at scale compared to other large scale training systems [22]. Although data parallelism or model parallelism approaches can be adopted for training, ML systems always feature repeating, high bandwidth communication patterns and a predictable workload. These requirements are an ideal fit for the scheduled topology shifts that the Apollo OCS platform supports, to achieve fast, flexible, and distributed training.

From a reliability perspective, link down events can cause large performance degradation for a Torus-style network [1, 5]. This places extra premium on reliability/availability of corresponding hardware, in comparison to application in a traditional datacenter network which achieves higher fault tolerance through link fanout and redundancy. Lastly, with limited buffering at the computing node, the tightly coupled communication for ML is similar to high performance computing, which requires designing optical transceivers for low latency (<100ns).

More details of our ML system will be described in a future publication.

All the above use cases utilizing Apollo OCS focus on networking inside the datacenters. These concepts can be extended to the network hierarchy beyond the datacenter network, i.e., intra-campus/inter-datacenter networking.

## 3. Optical Circuit Switching Technology

Optical circuit switching for communications has been an area of intense investigation and development, producing a rich variety of optical devices with different physical mechanisms to implement the switching function [20, 23-25]. Spurred on by the telecom bubble, a significant amount of development and commercial investment in OCS technologies occurred during the late 1990s to early 2000s [20, 23, 24]. For the initial telecom applications, smaller scale switches were used for protection switching while larger scale optical cross connects provided fast bandwidth provisioning and network management [24, 25].

| Technology | Relative Cost* | Port Count | Switching Time | Insertion Loss** | Driving Voltage (volts) | Latching |
|---|---|---|---|---|---|---|
| MEMS [25, 26] | Medium | 320x320 | ms | <3dB | 100s | No |
| Robotic [27] | Medium | 1008x1008 | mins*** | <1dB | | Yes |
| Piezo [28] | High | 384x384 | ms | <2.5dB | 10s | No |
| Guided Wave [29] | Low | 16x16 | ms | <6dB | 1s | No |
| Wavelength Switching [30] | TBD | 100x100 | ns | <6dB | 0 | Yes |

*Based on scale indicated. **Includes connector losses. ***Per connection

**Table 1: Cost, scale, performance, and reliability/availability comparison of various OCS technologies. Costs and insertion loss are based on the port count indicated. For reliability/availability, the high voltage operation required for MEMS-based systems can be limiting if quality control is insufficient. Latching operation is also desirable, but can be compensated through infrastructure-level power redundancy.**

Table 1 compares cost, scale, performance, and reliability/availability of various OCS technologies. Systems employing piezo-electric actuation, robotics to mechanically reconfigure a patch panel, and MEMS are among those previously achieving some limited commercial adoption. In terms of scaling to the large number of port counts required for datacenter applications at acceptable costs, MEMS-based systems had demonstrated the most promise, with the realization of systems achieving greater than 1000x1000 interconnectivity [25]. The robotics configured OCS, although able to scale to larger port counts while supporting any fiber type (single and multi mode fibers), suffers from slow switching speeds that are further compounded by the need to serialize switching of connections. Piezo-based systems have higher costs due to assembly complexity and guided wave switching has limited scale with high losses, although offering the lowest costs and size due to integration. Wavelength-switching-based schemes have been extensively investigated by the research community, primarily for faster optical switching (optical packet switching (OPS)/optical burst switching (OBS)). However, even for slower switching applications, wavelength switching lacks future proofing, as the channel spacing and width of the arrayed waveguide grating (AWG) limits the link speed and wavelength plan.

MEMS-based OCS technology holds a number of benefits relative to EPSes, being data rate and wavelength agnostic, low latency, and extremely energy efficient.

*Data rate and wavelength agnostic*: The MEMS-based OCS simply deflects light from the input port to the desired output port, normally using two arrays of mirrors which can be tilted about two axes (steered in a three dimensional fashion), as shown in Figure 4. Due to the broadband, passive nature of the optical path, increasing the line rate as well as the number of per-port multiplexed wavelengths can be achieved with the same OCS hardware, allowing reuse of the OCS over multiple generations of optical transceiver technologies with the only limit being the capacity/bandwidth of the single mode fiber and associated components interfacing to the mirrors. Analogous to power and cooling infrastructure, the corresponding optical fabric thus becomes part of the datacenter facility, with associated CapEx costs amortized over the lifetime of the building (multi decade versus multi year).

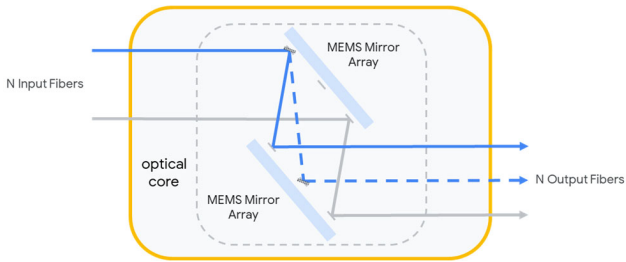

**Figure 4: Principle of 3D MEMS-based, NxN optical switch. Mirrors are tilted about two axes to steer optical beams from one path (solid line) to another path (dashed line). Each path contains two mirrors and two input/output fibers or ports. For simplicity, the figure illustrates rotation along only one axis.**

*Low power consumption*: Since there is no per-packet processing, per-bit energy consumption can be orders of magnitude lower than their EPS counterparts. In addition, the mirrors are electrically equivalent to a capacitive load. Although the voltages required to tune the MEMS mirrors can be large (one to a few hundred volts), the power consumption required to maintain mirror locations can thus be extremely low. A well designed high voltage driver circuit can consume on the order of tens of milliwatts of power per mirror and port.

*Low latency*: Again, since there is no per-packet processing, only a small amount of latency is added. Latency is set by speed-of-light propagation delay, i.e., ~5ns per meter in optical fiber and ~3.3ns per meter in free space. In comparison, an equivalent throughput EPS would add several tens if not hundreds of nanoseconds of delay per network hop. This gap becomes more critical for latency sensitive ML applications.

Despite these favorable characteristics, datacenter economics, scale, and performance requirements impose a number of requirements and challenges for OCS hardware:

*Larger scale*: The largest commercial OCS we were aware of at the time supported a few tens to hundreds of duplex ports. For integration into datacenters, OCS scaling to at least hundreds of ports is desirable to support connectivity to a sufficient number of ABs (i.e., large fabric size through scale out). More importantly, the demand for optical switching from traditional telecom applications is insufficient to support the volumes required by datacenter applications. Tens to hundreds of switches must be manufactured and tested per week. This requires not only increase in manufacturing assembly and test capacity, but corresponding design for manufacturability and high reliability/quality at scale (i.e., stringent Telcordia testing and development of reliability models through appropriate aging tests). Analogous to the transformation needed for optical transceivers manufactured for telecom versus datacom, this scaling challenge is a critical gap not addressed by the optics industry.

*Faster switching time*: Commercial OCS switching times are typically between 10-20ms, limited by OCS control software and mirror configuration time. Receiver initialization of the optical transceiver also constrains the use of faster switching (i.e., requires burst mode operation [31]). For initial datacenter use cases we described above, these millisecond order switching times are sufficient. However, we believe there are significant, future opportunities for large-scale OCS supporting microsecond-order switching times (discussed in Section 6).

*Lower insertion and return loss*: Optical link budget is a very precious commodity for datacenter network fabrics, greatly affecting cost and power consumption of the optical transceiver technology. Use of cost-effective, low-power optical transceivers with moderate link power budget requires driving down the insertion loss through the OCS, ideally below 2dB. The requirement for low return loss stems from the application of bidirectional links, described in detail in Section 4.

*Wide wavelength range operation*: When scaling to beyond multi-Tb/s, datacenter optical interconnect may need to employ large channel count WDM to minimize cost and deployment challenges, thus requiring operation across multiple communication bands (O, S, C, L). Sustaining aggressive return loss performance becomes challenging across such a wide band of operation.

*Non-blocking*: Although some switch hardware implementations may employ re-arrangeably non-blocking schemes, a strictly non-blocking switch is highly desirable to maximize flexibility, minimize disruption to the network, as well as reduce the burden on controls and software.

*High availability*: With the large amount of networking traffic transiting the OCS, high reliability and availability are essential for robust network performance. Overall availability may be increased with the addition of spare ports, by designing in redundancy (power, fans), enabling higher failure rate components to be field replaceable, and latching operation (i.e., last known state preserved under loss-of-power events).

*Lower cost*: The cost of integrated MEMS-based OCS has traditionally been a barrier to entry in the datacenter. At the same time, the underlying chip technology is inherently inexpensive due to fabrication within a silicon wafer process. Just as the large market within the datacenter drove cost down for Ethernet switch ASICs, a demand shift from telecom to datacom volumes can equivalently drive costs down for MEMS-based OCS modules and chips. Overall, we set a budget of 15% of network costs to be in support of optical switching. We did so because we knew from our own studies and the research literature that the benefits could be >2x that and we required at least that level of benefit to embark on a challenging path to productionizing OCS.

As there are fixed, inherent tradeoffs among port count, switching time, and insertion loss of current OCS designs, exploration of alternative architectures continues to be a promising area of work for delivering significant impact.

## 4. Hardware Components of Apollo

We developed three critical hardware components, the OCS, WDM transceivers, and optical circulators, to realize a cost-effective, large-scale optical switching layer.

### 4.1 Palomar OCS

For the first several years of deployment, a vendor-based OCS solution was adopted in the Apollo layer. However, due to the difficulties in maintaining reliability and quality of this solution at scale, the decision was made to internally develop an OCS system.

Figure 5 shows the high-level optical design and operation principles of the Palomar OCS. The input/output optical signals enter the optical core through two-dimensional (2D) fiber collimator arrays. Each collimator array consists of an NxN fiber array and 2D lens array. The optical core consists of two sets of 2D MEMS mirror arrays. Each inband optical signal traverses through a port in each collimator array and two MEMS mirrors, as indicated by the green line of Figure 5. Mirrors are actuated and tilted to switch the signal to a corresponding input/output collimator fiber. The entire end-to-end optical path is broadband and reciprocal, for data rate agnostic and bidirectional communication across the OCS. Superposed with the inband signal path, a monitoring channel (red arrows, Figure 5) assists with tuning of the mirrors. Each MEMS array is injected with 850nm light. The reflected monitor signals are received at a matching camera module. A servo (control hardware/firmware) utilizes the camera image feedback to optimize MEMS actuation for minimum loss of the optical signal path. A pair of injection/camera modules controls each 2D MEMS array.

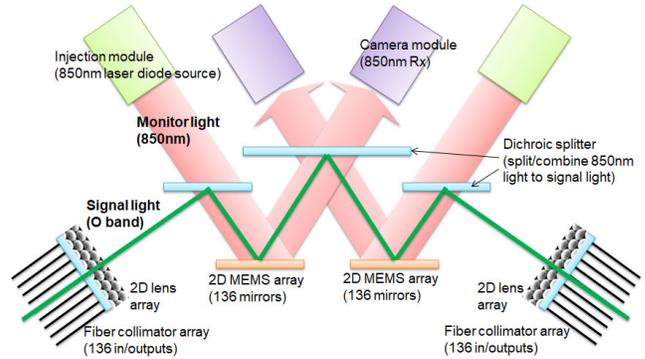

**Figure 5: Illustration of design and optical path of Palomar OCS optical core. Inband optical signal path indicated by green line. Superposed with the inband signal path, a monitoring channel at the 850nm wavelength (red arrows) assists with tuning of the mirrors.**

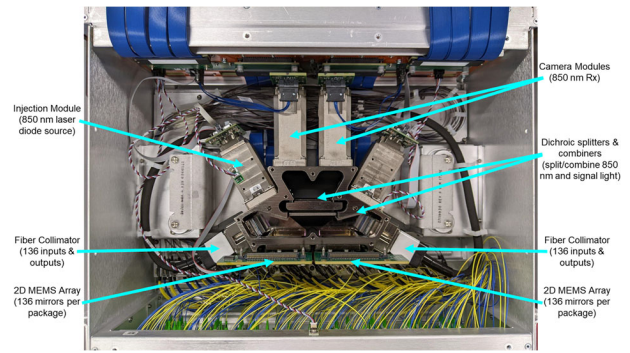

**Figure 6: Photograph of Palomar OCS optical core (with top cover removed to see the interior), and corresponding key components: a) fiber collimators, b) camera modules, c) packaged MEMS, d) injection modules, and e) dichroic splitter and combiners.**

By implementing mirror controls based on image processing of a single camera image per MEMS array, the control scheme is significantly simplified in comparison to conventional approaches which require individual monitoring and/or photodetector hardware per mirror. This design choice was critical to realizing a low-cost, manufacturable OCS solution. The above design yields a non-blocking, 136x136 OCS with bijective, any-to-any input to output port connectivity. Figure 6 shows the optical core and corresponding key components.

At the heart of each optical core are the MEMS micro-mirrors that perform the optical steering function. In order to provide any in to any out beam steering functionality, two MEMS mirror packages are used to provide four degrees of freedom. This is necessary to optimally couple the light from the inputs to the outputs. Figure 7a) shows a photograph of one of the MEMS mirror packages. Within each ceramic package is a large MEMS die that has 176 individually controllable mirrors that are eventually down-selected to yield the final 136 mirrors in the fully calibrated systems. On top of the ceramic package, a window is placed to seal the MEMS die and protect the micro-mirrors from dust and humidity over the lifetime of the OCS product. An angled window is used to prevent zero-volt state coupling and potential back reflections to the fiber collimator arrays. The window coating is designed to allow both the monitor wavelength (850 nm) and the data wavelengths to pass through with minimal loss. This is extremely important because each data path has eight total passes through the window coatings (top and bottom sides). Shown in Figure 7b), each micro-mirror features a highly reflective gold coating to minimize optical loss. High-voltage (HV) driver signals are brought to the mirror control axes through four wirebonds per mirror (one per axis). The MEMS mirrors are manufactured using a DRIE (deep reactive ion etching)-based process to produce large diameter, flat micro-mirrors with high reflectivity.

the other two left at zero (fairly standard for 3D mirrors [24]). After the initial state is set, the beam position on the camera is then moved to the optimum position, as described earlier. Figure 8b) shows a rear view of the OCS chassis. All of the PCBAs in the back of the chassis are field replaceable units (FRUs). The power supplies and fan modules operate in a 1+1 and 2+2 redundancy mode, respectively, and can be hot swapped while maintaining functionality. Although the mirror state cannot be maintained when HV driver boards are hot swapped, designing them to be a FRU was critical as high voltage ICs and operation present one of the largest reliability challenges for MEMS-based optical switching systems. The maximum power consumption of the entire system is 108W, which is a fraction of the power of an EPS system with the same switch capacity. With the above hardware architecture, tens of thousands of 136x136 duplex port OCS (eight spare ports) were manufactured and deployed over the past decade.

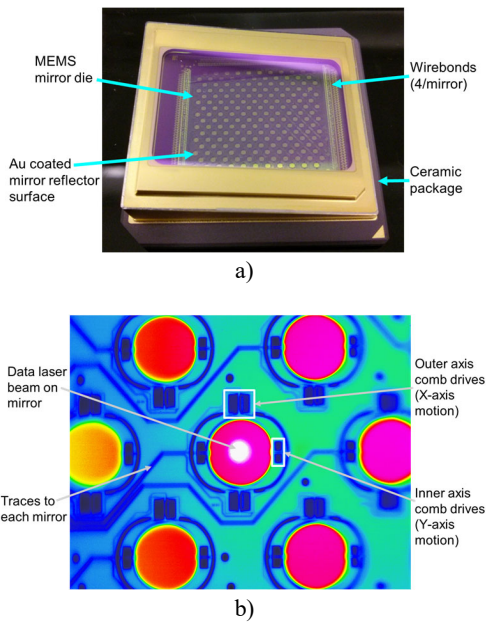

Figure 7: a) Photograph of a Palomar MEMS mirror package. Inside each ceramic package is a single large die with 176 individually controllable micro-mirrors. b) Thermal image of MEMS mirrors. Each mirror has four comb drive regions to rotate it in two directions.

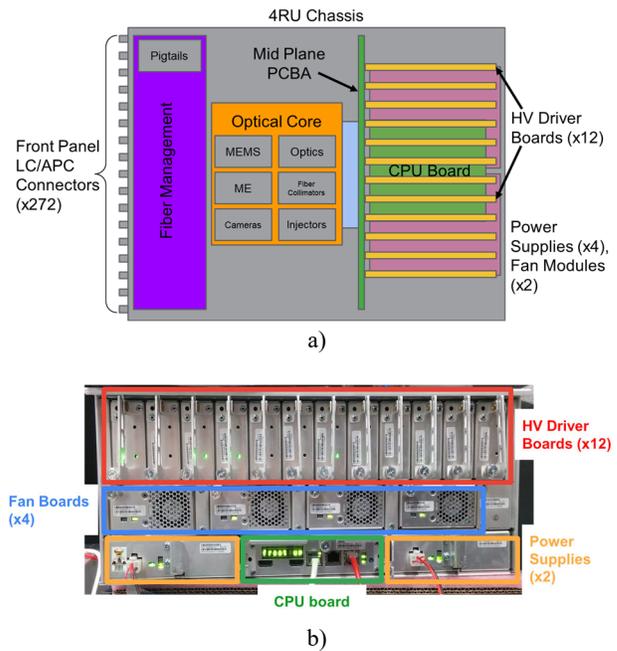

Figure 8: a) System level diagram of Palomar OCS showing the two main sections: front half with fiber management and the optical core. The back chassis with the primary CPU board, power supplies, fan modules, and HV driver boards. b) Rear view of the Palomar OCS back chassis showing the FRUs.

Figure 8a) shows the system level diagram of the OCS chassis. The CPU receives port connection commands from the user, which are then sent to an FPGA connected to a set of HV driver boards. The driver board outputs the initial-state mirror voltages, generated with digital-to-analog converter (DAC) and amplifier ICs. Each mirror has four voltage inputs, two of which are driven to actuate the mirror,

For the on-device software, Palomar runs Linux, utilizing the same software stack and base OS as our other datacenter networking devices (i.e., EPSes). The commonality in management plane interfaces allows strong leverage of infrastructure and seamless integration with existing datacenter workflows. Portions of the codebase that dealt with managing EPSes were replaced with image processing algorithms and state machines tracking not only mirror position, but also camera and video pipeline health, injector brightness, and other critical system parameters. Hardware offloads are also utilized to enable smooth mirror control and improved video performance with a minimal amount of CPU involvement. Significant development effort was invested in improving telemetry and anomaly reporting to account for the complexity of the hardware and the software interactions that manage it.

manufacturability. To enable high throughput and yield, most of the product's opto-mechanical tolerances are pushed to the MEMS mirrors. This also enables rapid assembly and easy alignment of the optical core. To produce the Palomar OCS, we developed custom testers, alignment, and assembly stations for the MEMS mirrors, fiber collimators, optical core and its constituent components, and the full OCS product. Each MEMS mirror is fully tested at the wafer foundry using a prober-based solution that can address each of the 176 mirrors on the die during a single touch-down event. A custom, automated alignment tool was developed to place each 2D lens array onto the face of the fiber collimators with sub-micron accuracy. Finally, each system must go through full MEMS mirror calibration for port selection. All available MEMS mirrors are optimized for all possible connection states. If both MEMS mirror arrays had 100% good mirrors, this would be a total of 176x176 possibilities (30,976). In reality, there are almost always less than 30k initial port combinations. Of these, the very best are then plugged into the OCS front panel for use. This final set of used MEMS mirrors provides a total of 136x136 possible connections (18,496). The voltages needed to optimally couple these 18,496 combinations are stored within each unit as a custom mapping for that particular OCS.

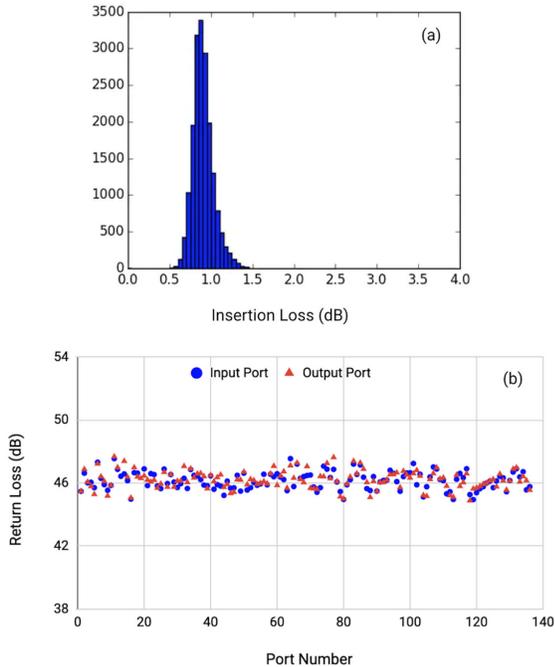

**Figure 9:** a) Representative Palomar OCS insertion loss histogram for 136x136 (18,496) cross-connections. b) Return loss versus port number for 136 input/output ports, with the OCS configured for 1:1 connections (input 1-output 1, input 2-output 2, …).

Figure 9 shows some representative insertion loss and return loss data for the Palomar OCS. Insertion losses are typically <2dB for all NxN permutations of connectivity. The tail in the distributions is nominally due to splice and connector loss variation. Return loss is typically -46dB, with a nominal spec of <-38dB. The major components of reflection come from the fiber collimators, at the interfaces between the fiber array and 2D lens array. The stringent return loss requirement stems from the use of bidirectional communication along each optical path, as any single reflection superposes directly on top of the main optical signal to degrade signal-to-noise ratio. Multilevel PAM-based communication further increases sensitivity to these reflections (discussed at the end of Section 4).

Manufacturing a custom OCS at scale requires that the product be designed from the start for yield, throughput, and

### 4.2 WDM Optical Transceivers

For telecom applications, WDM technology was integral to scaling the capacity of high-value long-haul fiber links by multiplexing a large number of wavelength channels per fiber. Similarly in the datacenter, as the bandwidth scales, adoption of WDM was critical to improve the cabling efficiency and support a large and ever-growing compute infrastructure. This pushed the need to optimize cost, performance, and develop an industry ecosystem befitting the datacom application.

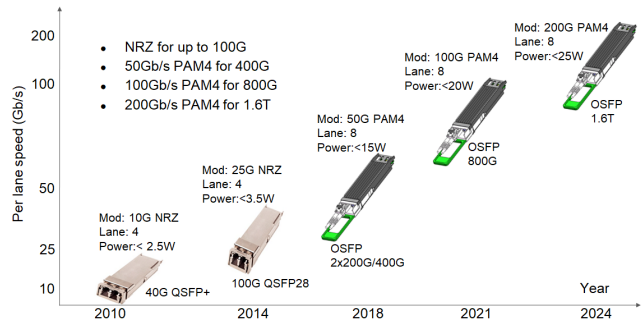

**Figure 10:** WDM single mode interconnect review and roadmap [32].

Figure 10 shows the corresponding WDM single mode interconnect roadmap developed over the past decade by Google [20, 32-34]. For the initial Apollo-based network, we chose the IEEE standard based 40Gb/s long reach (LR4)/single mode CWDM4 solution as the baseline for our inter-AB interconnect design. To support the higher loss budget due to the OCS and circulators, transceiver design emphasized low optical component (wavelength mux/demux) and packaging losses (superior opto-mechanical design). In addition, the presence of directly modulated laser (DML) technology was fortuitous for this requirement, due to the high power nature of DMLs.

Thereafter, a variety of technology directions and trends are notable. First, in order to support direct interop between various generation ABs, it was critical to maintain the same CWDM4 wavelength grid. This necessitated development of key, critical component technologies starting at the 25G per optical lane generation well in advance, most importantly uncooled CWDM DMLs. This eventually led to after-the-fact creation of the CWDM4 MSA [35].

Second, in order to keep scaling down the cost, power, and density per Gb/s and enable the use of the switch ASIC/Moore's Law improvements [20, 36], continuous speed up of each lane was essential. This was achieved through development of a variety of key technologies: optical, electrical, and signal processing. Figure 11c) highlights these three key areas which initially diverged away from standard LR technology. For the optics, we worked with the industry to develop faster optical components (lasers/photo-detectors), migrating from DMLs to more recently the use of externally modulated lasers (EMLs), due to higher speed and extinction ratio requirements (critical for mitigating multi-path interference (MPI) effects enhanced by bidirectional communication). For high-speed IC/electrical technologies, we migrated from analog-based clock-and-data recovery (CDR) solutions to digital signal processing (DSP)-based ASICs [37]. The DSP ASIC provided a more robust, scalable solution by relaxing the requirement on optical and analog electrical components at the expense of increased power consumption and latency. In addition, we leveraged the new found digital capabilities to develop algorithms that mitigate MPI impairments inherent in bidirectional links [38-40]. This also includes the development of forward error correction (FEC) techniques in order to support the higher link budgets needed for the Apollo-based links.

One additional and crucial element of the interconnect development was the requirement for backward compatibility of the optical transceiver technology. In addition to wavelength grid compatibility, this meant the optical performance specifications and high speed data path needed to support various line rates. The current generation transceiver must also support a superset of all transmitter and receiver dynamic ranges of the previous generation transceivers. In particular for the receiver, supporting a range of sensitivity performance across various line rates becomes challenging, due to constraints on the trans-impedance amplifier (TIA) design: a lower data rate requires higher trans-impedance gain to support low received power, while a higher data rate requires lower trans-impedance gain to support larger bandwidth.

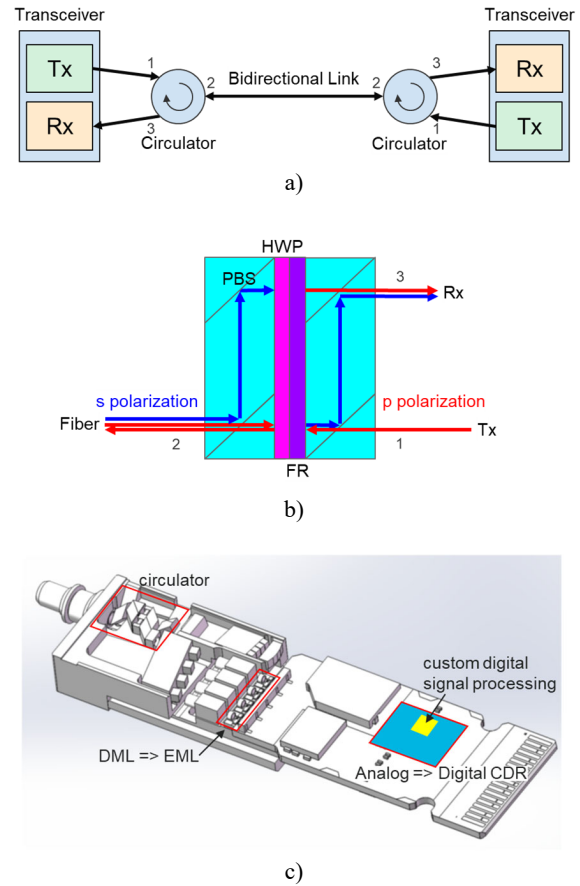

Figure 11: a) The optical circulator is a three-port non-reciprocal device that has a cyclic connectivity. Input into port 1 is directed to port 2, input into port 2 is directed to port 3. The circulator thus converts a traditional duplex optical transceiver into a bi-directional one. b) Example integrated circulator implementation. Blue lines indicate s-polarization, red lines indicate p-polarization. PBS - polarizing beam splitter, FR - faraday rotator, HWP - half wave plate. Numbers indicate corresponding port numbers matching a). c) bidi CWDM4 optical transceiver with integrated circulator.

### 4.3 Optical Circulators

As stated earlier, the optical circulator enables bidirectional operation of Apollo links. Figure 11a) shows the use of optical circulators in a bi-directional link which reduces both the number of OCS ports and fiber cables required by half. The optical circulator is a three-port non-reciprocal device that has a cyclic connectivity. Input into port 1 is directed to port 2, input into port 2 is directed to port 3. This functionality can be realized through the use of birefringent crystals, magneto-optical Faraday rotators (typically made of Garnet), and polarizers [41]. Figure 11b) illustrates an example implementation. Although the optical signals running in opposing directions do superpose on top of each other along the fiber and OCS, they do not directly interact with each other due to the bosonic nature of photons.

Before we drove high volume use in the datacenter, the circulator was commonly employed in erbium-doped-fiber amplifiers, allowing a double pass through the doped fiber with a circulator on one end and a reflector on the other to increase gain. Application was mostly in the C-band wavelength range with fairly limited volumes. Use of proper optical coatings and optical re-design allowed extension of circulator capability to O-band/CWDM4 wavelength operation. Optimizations to reduce return loss as well as enhance directivity (i.e., minimizing port 1 to 3 crosstalk) were critical, as corresponding stray light is effectively equivalent to having a reflection in the link. Similar to the OCS, the broadband nature of the circulator allows its reuse across multiple generations of CWDM4-based optical transceiver technologies, with costs amortized accordingly.

Although initial implementations of the circulator were external to the optical transceivers (chassis containing circulators at top of rack), the circulator can also be integrated into the transceiver for further performance, size, and cost reduction at the expense of the ability to reuse across different generation transceivers. Figure 11c) shows an example implementation of circulator transceiver integration. Removing connector losses through integration decreases overall link insertion loss while increasing density.

### 4.4 Link Modeling and Design

OCS plus circulator-based bidirectional links impose a number of unique physical impairments, most notably higher insertion losses and MPI effects. In order to support such links at scale, link design must bring together performance parameters of all critical components (bandwidth, noise, dispersion, transmit power/receiver responsivity, insertion/return losses, extinction ratios, etc.). As an example of the required holistic link design, Figure 12 shows the link model a) and corresponding receiver sensitivity results b) simulated for 50Gb/s PAM4. In this result, we highlight the sensitivity to multiple sources of return loss/reflections and crosstalk along the link, primarily from the OCS and circulators. The substantial link penalties with higher MPI clearly demonstrates the need for tight specification of the OCS, circulators, as well as the fiber plant itself.

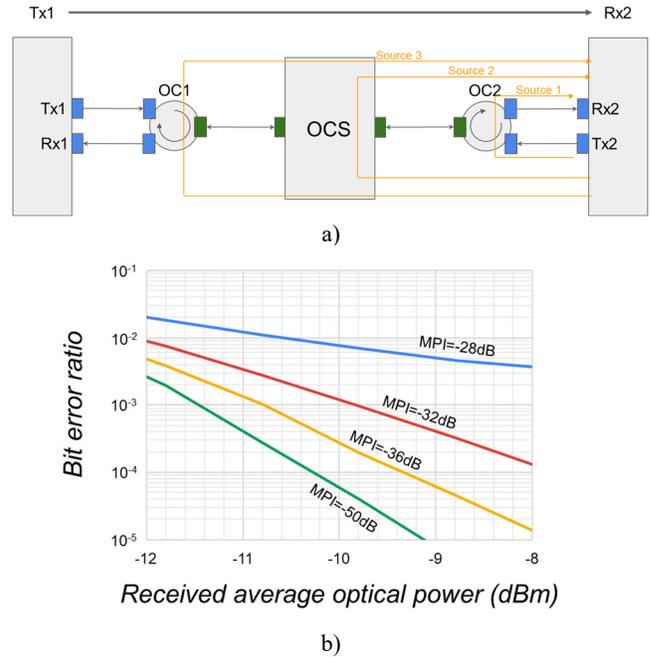

**Figure 12: a) Apollo link model including OCS, circulators, and corresponding sources of reflection. b) Simulated receiver sensitivity results for 50Gb/s PAM4 transmission.**

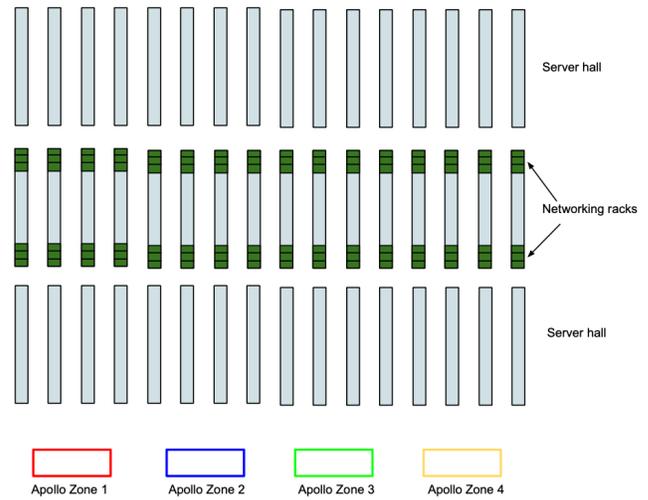

**Figure 13: Apollo physical layout on the datacenter floor.**

### 5. Apollo Physical Layout

The physical layout of Apollo on the datacenter floor is shown in Figure 13. The maximum OCS count of 256 is split across four separate locations or zones with sufficient separation in order to distribute the physical failure domain. Each Apollo Zone supports eight racks capable of supporting eight OCSes per rack. Figure 14a) shows half of an Apollo Zone, with eight OCSes populating each rack. On the other end, ABs are located in the middle of the datacenter floor (Networking racks, Figure 13).

Links running to/from each AB to OCS have a maximum length of several hundred meters. With these modest reaches for WDM/LR technologies, optical loss and link budget have dominated link design thus far (versus dispersion and reach limited design). Home run fibers (i.e., no patch panels) with APC connectors mate directly with the OCS and circulator chassis to minimize insertion and return losses. The lack of constraint on reach allows more flexible placement of the various networking elements, to facilitate planning and layout.

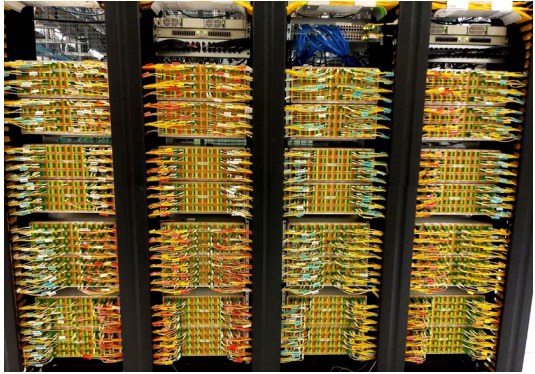

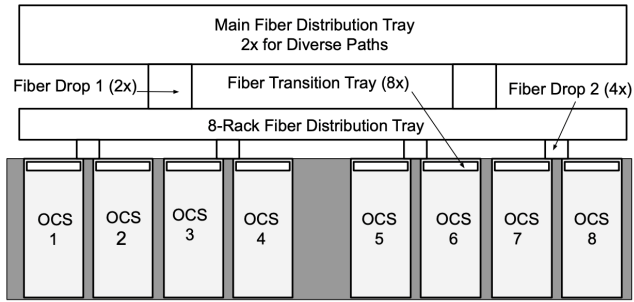

a)

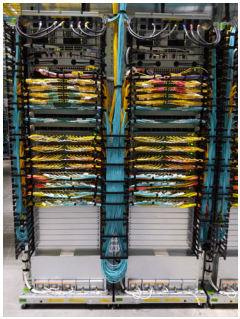
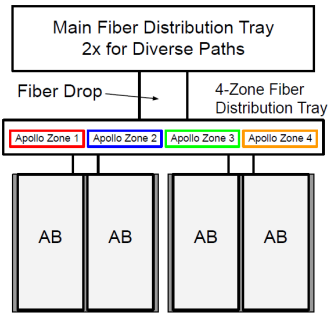

b)

**Figure 14:** a) Half of an Apollo Zone (top) and fiber cable distribution to OCS racks (bottom). b) Half of an AB housed within two, conjoined racks (left) and fiber cable distribution to networking racks (right).

To deal with the large number of fiber connections, on the OCS side, there are three layers of fiber distribution with two fiber drops (Figure 14a)). Main distribution trays (2x) are separated by six feet for physical diversity and feed fibers to all four Apollo zones. Another distribution tray at a lower elevation feeds fibers to all eight OCS racks. At the top of each OCS rack, there is a fiber distribution tray managing fiber connections to OCS ports. The main fiber drop occurs above the eight rack distribution tray, and fiber drop 2 is centered above a pair of OCS racks. On the other end, each AB is split into four racks. Two racks are conjoined (i.e., half of an AB is shown in Figure 14b)) and placed in adjacent aisles. Main distribution trays (2x) feed fibers to four fiber distribution trays above the networking racks, one for each Apollo Zone.

In addition to physical diversity, power redundancy is achieved through dual power feeds from circuits A/B with eight outlets for each Apollo Zone, and two outlets for a pair of OCS racks. 60kW uninterruptible power supply (UPS) with 30kW per feed is reserved per fabric.

## 6. Future Work

Immediate future work would be to expand the above concepts and capability of an optical switching layer to other areas of the network hierarchy, specifically intra-campus and inter-datacenter networking. Longer term work includes extensibility to different layers of the datacenter network as well as further development of ML and high performance computing applications, which are a natural fit for circuit switched technology due to more predictable job scheduling and/or credit-based networking.

In order to keep up with these new as well as growth of existing network requirements, future hardware evolution of optical technologies include the following: 1) Larger port count OCS to enable further scale out with increased striping and to increase efficiency through more flexible topology engineering. 2) Faster switching speed and/or smaller radix but lower cost OCS to allow adoption in lower layers of the datacenter network for shorter, more bursty traffic flows (i.e., TOR to AB traffic) or flexible bandwidth provisioning with adjustment of TOR oversub ratio. 3) Further improvements in reliability and availability for larger OCS/failure domains and/or more uptime sensitive applications. 4) Lower insertion and return loss for extensibility of the optical interconnect roadmap for continued low power consumption and cost of the transceivers. In addition to facilitating transceiver roadmap, the last point expands application to more novel use cases such as cascading of multiple OCSes into a Clos configuration for a larger size optical fabric, as outlined in [42].

In terms of specific technologies for implementation, we believe MEMS-based switching has further runway on all axes (scale, speed, loss). Piezo-based switching has some fundamental advantages in insertion and return loss, which may tip the scales to this technology at a given switch port count if yields and reliability for MEMS-based solutions are no longer manageable. For faster switching, a wide variety of technology approaches have been researched extensively: 1) MEMS [43, 44], 2) waveguide based [20, 45], and 3) wavelength-based switching with AWGs [20, 30, 46-48]. All must take into account the interconnect roadmap, as this closely dictates the optimal solution from a holistic, system perspective. The use of coherent technologies for shorter reaches in the datacenter [40, 49] may support some of the traditional burdens of wavelength switching, which requires a fast, wavelength-tuning function from the transceiver [50, 51]. However, additional myriad challenges and requirements remain for this and other implementations of fast switching (burst mode receivers, suppressing crosstalk in switch fabric, high optical losses, manufacturability/cost).

## 7. Related Work

As reviewed in Section 3, various MEMS-based OCS systems have been investigated and manufactured by the industry [24-26]. These include larger port count systems [25] than this work. However, the use of camera image processing for mirror controls and fulfilling datacenter volumes/requirements over a decade make the Palomar OCS unique from these other solutions. The use of circulators for increasing OCS and other optical switch radix has been proposed in other work [15]. However, these are at best, small scale test bed or prototype implementations, and not delivered at millions of port count scales as was done over the past decade with Apollo. WDM transceivers are part of the IEEE 802.3 standards. In truth, our development of WDM transceivers has led and defined the direction of the WDM optics industry roadmap for the past decade. In addition, the optimization of transceiver designs for Apollo OCS and circulator based bidirectional links that are backward compatible is new and a unique contribution of our work.

At the system level, there have been multiple instantiations of OCS-based datacenter network proposals [12-20]. The concept of topology engineering is prevalent through many of these works. The concepts of fabric expansion and rapid tech refresh is a unique contribution of this work. Finally, all of the above related works are proof-of-concept, testbed scale demonstrations, in stark contrast to the hyperscale deployment achieved with Apollo and associated challenges overcome over a decade, across a real-world cloud infrastructure.

A large body of work also exists for finer granularity OPS and OBS with the goal of achieving switching in the optical domain on nanosecond and microsecond-order timescales, respectively [20, 30]. For this paper, we have focused our discussion on OCS due to the still limited technology maturity of OPS/OBS. In particular, the lack of efficient and simple buffering in the optical domain and need for high-speed burst mode receivers and/or clocking schemes that circumvent such requirements remain significant barriers to the adoption of OPS and OBS.

## 8. Conclusion

In this paper, we motivated the use of optical switching for datacenter networking with our previous work on datacenter networks and industry-wide proposals outlining a variety of use cases and the advantages of optical switching for such applications. We then presented our implementation of Apollo: an optical switching layer for datacenter networks at scale, comprising key hardware components of the OCS, WDM optical transceiver technology, and circulators. These capabilities combined have allowed us to achieve: 1) the lowest cost networks at scale, with pay-as-you-grow capability, 2) fastest adoption of the latest generation optics and networking that we develop, and 3) dynamic network reconfiguration capabilities for highest efficiency and/or performance.

In closing, we believe this work is just the initial step in moving the entire datacenter architecture into the optical domain, with a plethora of high impact future work to follow.

## 8. Acknowledgements

The authors would like to acknowledge the Apollo and Palomar teams, Platforms, Net Infra, Google Global Networking (GGN), Global Capacity Delivery (GCD), Datacenter Operations, Site Reliability Engineering (SRE), Program Management Organization (PMO), and industry/vendor partners for making Apollo a reality at scale. We would also like to thank Parthasarathy Ranganathan, Lieven Verslegers, and Shuang Yin for review of the manuscript.## REFERENCES

[1] L. A. Barroso, U. Hölzle, and P. Ranganathan. *The datacenter as a computer: Designing warehouse-scale machines, third edition*. Morgan and Claypool Publishers, 2018.

[2] J. Jumper, R. Evans, A. Pritzel, T. Green, M. Figurnov, O. Ronneberger, K. Tunyasuvunakool, R. Bates, A. Žídek, A. Potapenko, A. Bridgland, C. Meyer, S. A. A. Kohl, A. J. Ballard, A. Cowie, B. Romera-Paredes, S. Nikolov, R. Jain, J. Adler, T. Back, S. Petersen, D. Reiman, E. Clancy, M. Zielinski, M. Steinegger, M. Pacholska, T. Berghammer, S. Bodenstein, D. Silver, O. Vinyals, A. W. Senior, K. Kavukcuoglu, P. Kohli, and D. Hassabis. Highly accurate protein structure prediction with AlphaFold. *Nature*, 596: 583-589, 2021.

[3] D. Silver, A. Huang, C. J. Maddison, A. Guez, L. Sifre, G. van den Driessche, J. Schrittwieser, I. Antonoglou, V. Panneershelvam, M. Lanctot, S. Dieleman, D. Grewe, J. Nham, N. Kalchbrenner, I. Sutskever, T. Lillicrap, M. Leach, K. Kavukcuoglu, T. Graepel, and D. Hassabis. Mastering the game of Go with deep neural networks and tree search. *Nature*, 529: 484-503, 2016.

[4] N. P. Jouppi, C. Young, N. Patil, D. Patterson, G. Agrawal, R. Bajwa, S. Bates, S. Bhatia, N. Boden, A. Borchers, R. Boyle, P. Cantin, C. Chao, C. Clark, J. Coriell, M. Daley, M. Dau, J. Dean, B. Gelb, T. V. Ghaemmaghami, R. Gottipati,12

13